\documentstyle[sprocl]{article}

\input{psfig.sty}

\def\NPB{{\em Nucl. Phys.} B}

\def\PRD{{\em Phys. Rev.} D}

\def\be{\begin{equation}}
\def\ee{\end{equation}}
\def\bea{\begin{eqnarray}}
\def\eea{\end{eqnarray}}

\begin{document}

\begin{flushright}                                
UMN-D-00-3 \\ July 2000 \\ \vspace{0.3in}
\end{flushright}

\title{NONPERTURBATIVE CALCULATION OF SCATTERING AMPLITUDES%
\footnote{To appear in the proceedings of               
the fourth workshop on Continuous Advances in QCD,
Minneapolis, Minnesota, May 12-14, 2000.}%
}

\author{J. R. Hiller%
\footnote{\baselineskip=14pt                            
Work supported in part by the Department of Energy,
contract DE-FG02-98ER41087.}%
}

\address{%
Department of Physics, 
University of Minnesota Duluth \\ 
Duluth, MN 55812, USA \\
E-mail: jhiller@d.umn.edu} 

\maketitle
\abstracts{%
A method for the nonperturbative calculation of scattering
amplitudes and cross sections is discussed in the context 
of light-cone quantization.  The Lanczos-based recursion
method of Haydock is suggested for the computation of
matrix elements of the resolvent for the light-cone Hamiltonian,
from which the $T$-matrix can be constructed.  The scattering
of composite particles is handled by a generalization of
a formulation given by Wick.}

\section{Introduction}

Recently there has been a significant amount of progress in
the nonperturbative calculation of bound-state properties
in light-cone quantized field theories.~\cite{review,PV}  Comparable
progress in the calculation of scattering states has been
lacking, although techniques have been proposed.~\cite{scattering}
The ability to compute outcomes of scattering processes is not
only of interest in its own right, but also useful for construction
of renormalization conditions in bound-state calculations.~\cite{ae}
Here we consider a calculational scheme based on numerical
approximations to resolvents~\cite{Haydock} and on a
generalization of a Hamiltonian formulation for the
scattering of composites.~\cite{Wick}

The construction is done in terms of light-cone coordinates,
where $t+z$ takes the role of time.~\cite{Dirac,review}  This
is not a necessary choice, but it does bring several advantages
and makes immediate contact with work that uses these coordinates
for bound-state problems.~\cite{review,PV}  The advantages
include a simple vacuum, well defined Fock-state expansions,
and kinematical boosts.  This last advantage is particularly important
for the scattering of composites because one can compute the
state of a composite particle in one frame and boost to other
frames without again referencing the interaction.

The remainder of this paper is structured as follows:
Section~\ref{sec:CrossSections} contains a description
of the notation used for light-cone coordinates and a
formulation of two-body cross sections in terms of
these coordinates.  The key quantity in this formulation
is a matrix element of a resolvent; a numerical method
for its computation is discussed in Sec.~\ref{sec:Recursion}
and applied to a nonrelativistic example in Sec.~\ref{sec:NonRel}\@.
An extension to the scattering of composites is given in
Sec.~\ref{sec:Composites}.  Finally, Sec.~\ref{sec:Conclude}
provides a summary and indications of future work.

\section{Cross Sections in Light-Cone Coordinates} \label{sec:CrossSections}

We use the following notation for light-cone coordinates:
\be
x^\pm=t\pm z\,,\;\;{\bf x}_\perp=(x,y)\,,
\ee
where $x^+$ is taken as the time variable.  Momentum
components are similarly defined
\be
p^\pm=E\pm p_z\,,\;\;{\bf p}_\perp=(p_x,p_y)\,.  
\ee
A dot product is then written as
\be 
p\cdot x=\frac{1}{2}(p^+x^-+p^-x^+)-{\bf p}_\perp\cdot{\bf x}_\perp\,,
\ee
so that $p^-$ is conjugate to $x^+$ and is recognized as the 
light-cone energy.  The other momentum components form the
light-cone analog of the three-momentum
\be
\underline{p}\equiv(p^+,{\bf p}_\perp)\,.
\ee
The evolution of a state is determined by the
Hamiltonian ${\cal P}^-$; however, the term
``light-cone Hamiltonian'' is also
used~\cite{PauliBrodsky} for the combination
\be  \label{eq:HLC}
H_{\rm LC}=P^+{\cal P}^-\,,
\ee
where $P^+$ is the total momentum.
The fundamental eigenvalue problem is
\be  \label{eq:EigenProb}
H_{\rm LC}|p\rangle=\frac{M^2+p_\perp^2}{p^+/P^+}|p\rangle\,,\;\; 
\underline{\cal P}|p\rangle=\underline{p}|p\rangle\,.
\ee

For two-body scattering $A+B\rightarrow C+D$, the center-of-mass
cross section can be written~\cite{Peskin}
\be
\frac{d\sigma}{d\Omega}_{\rm cm} =\frac{1}{2E_A 2E_B v_{\rm rel}}
          \frac{|\vec{p}_C||{\cal M}_{fi}|^2}{16\pi^2 E_{\rm cm}}\,, 
\ee
where $M_{fi}$ is the invariant amplitude extracted from the
$S$ matrix
\be 
S_{fi}=\langle f|i\rangle+(2\pi)^4 \delta^{(4)}(p_f-p_i)i{\cal M}_{fi}\,, 
\ee
with states normalized according to
\be 
\langle f|i\rangle=(16\pi^3)^2p_A^+\delta(\underline{p}_A-\underline{p}_C)
                         p_B^+\delta(\underline{p}_B-\underline{p}_D)\,. 
\ee
The flux factors are determined by the current matrix element
\be 
\langle A|J^0|A\rangle=\frac{1}{2}\langle A|(J^++J^-)|A\rangle=2E_A\,,
\ee
with $E_A$ the energy of particle $A$.
For fermions, the spinors are normalized by $\bar{u}u=2m$.
From the optical theorem we also have
\be 
\sigma=\frac{2\,{\cal I}m {\cal M}_{ii}}{2 E_A 2 E_B v_{\rm rel}}\,.
\ee

In the light-cone interaction picture~\cite{Fuda} the invariant
amplitude can be related to a matrix element of a $T$ operator.
The $S$ matrix can be written as
\be  
S_{fi}=\langle f|U(x^+,x_0^+)|i\rangle=
           \langle f|i\rangle-2\pi i T_{fi}\delta(p_f^- - p_i^-)\,,  
\ee
with the $T$ operator defined by
\be  
T=V+V\frac{1}{p^-+i\epsilon-{\cal P}^-}V\,, 
\ee
where
\be
{\cal P}^-={\cal P}_0^- + V\,.
\ee
On removal of the momentum conserving delta function
\be 
T_{fi}=t_{fi}\delta(\underline{p}_f-\underline{p}_i)\,,
\ee  
we obtain
\be
{\cal M}_{fi}=-\frac{1}{16\pi^3}t_{fi}\,.      
\ee
We then need a means to compute $t_{fi}$.  In particular,
we need to handle matrix elements of the second term
in $T$ which involve the resolvent $(p^-+i\epsilon-{\cal P}^-)^{-1}$. 

\section{Recursion Method for Resolvents} \label{sec:Recursion}

We divide the matrix element of the $T$ operator into two pieces,
the Born term and the resolvent term
\be 
T_{fi}=V_{fi}+T_{fi}^{(2)}\,,\;\;
   T_{fi}^{(2)}\equiv \langle f|V\frac{1}{p^-+i\epsilon-{\cal P}^-}V|i\rangle\,. 
\ee
To the latter we apply the recursion method developed by Haydock~\cite{Haydock}
in the context of condensed matter physics.  This method is based on
the Lanczos diagonalization algorithm~\cite{Lanczos} which converts
the original Hamiltonian operator into an approximate tridiagonal 
representation.  The tridiagonal form is easily inverted to 
yield the resolvent.   

Define an initial vector $|u_1\rangle=cV|i\rangle$,
with $c$ a normalization constant,
and carry out Lanczos recursions~\cite{Lanczos}
\bea 
b_1&=&0 \,,\;\;
a_n=\langle u_n|{\cal P}^-|u_n\rangle\,,\;\; 
\langle u_n|u_n\rangle=1\,, 
\nonumber \\
b_{n+1}|u_{n+1}\rangle&=&{\cal P}^-|u_{n+1}\rangle
              -a_n|u_n\rangle-b_n|u_{n-1}\rangle\,. 
\eea
Although this could in principle be done analytically,
the usual approach is to begin with a large but sparse matrix
representation for ${\cal P}^-$ in a Fock basis.  A standard
way of constructing such a representation is discrete
light-cone quantization (DLCQ).~\cite{PauliBrodsky,review}
The vectors $|u_n\rangle$ provide an orthonormal basis in which
${\cal P}^-$ has the tridiagonal representation
\be 
{\cal P}^- \rightarrow \left(\begin{array}{llllll}
              a_1 & b_2 & 0 & 0 & 0 & \ldots \\
              b_2 & a_2 & b_3 & 0 & 0 & \ldots \\
                0 & b_3 & a_3 & b_4 & 0 & \ldots \\
                0 & 0 & b_4 & .  & . & \ldots \\
                0 & 0 & 0 & . & . & \ldots \\
                . & . & . & . & . & \ldots \end{array} \right)\equiv A\,, 
\ee
and the forward matrix element of the resolvent is approximated by the 
upper left matrix element of the inverse
\be 
T_{ii}^{(2)}\simeq 
   \frac{1}{c^2}\left\{[(p_i^-+i\epsilon) I-A]^{-1}\right\}_{1,1}\,. 
\ee
Non-forward matrix elements can be obtained by using the
block Lanczos procedure in place of the ordinary one
described above.~\cite{BlockLanczos}

Unfortunately, the discrete nature of the calculation
does not allow this finite matrix representation to
be useful in all respects.  In the limit of $\epsilon\rightarrow 0$,
the imaginary part of this approximation to $T_{ii}^{(2)}$
will contain only delta functions at the discrete eigenvalues
of the finite matrix approximation $A$.  To properly model
the contribution from the continuum, an extension must be 
made.  This can be done in terms of a continued fraction 
representation
\be 
c^2 T_{ii}^{(2)}=\frac{1}{p_i^-+i\epsilon-a_1-
                     \frac{b_2^2}{p_i^-+i\epsilon-a_2-
                       \frac{b_3^2}{p_i^-+i\epsilon-a_3-\cdots}}}\,, 
\ee
by replacing the last level with a {\em terminator}, a model 
for an infinite fraction.~\cite{HaydockNex}

A model can be generated in various ways; however, a
logical choice in the present situation is to use
the second Born approximation as a generator.  The 
matrix element of the resolvent for the free Hamiltonian
\be 
\langle i|V\frac{1}{p^-+i\epsilon-{\cal P}_0^-}V|i\rangle  
\ee
is computed exactly and then in continued fraction form
by iterating with ${\cal P}_0^-$ an equal number of times.
The exact form dictates the structure of the remaining
infinite fraction, which is used as a terminator for
the full resolvent.  

\section{Nonrelativistic Example} \label{sec:NonRel}

As an example of how the terminated recursion method works,
we consider nonrelativistic S-wave scattering from
an exponential potential.  This is a common choice for
testing numerical methods for scattering problems 
because it has an exact solution.

The partial wave expansion for the $T$-matrix is
\be 
\langle \vec{k}^{\,\prime}|T(E)|\vec{k}\rangle=
     \frac{1}{2\pi^2}\sum_l(2l+1)
                   T_l(k',k;E)P_l(\hat{k}^{\,\prime}\cdot\hat{k})\,, 
\ee
where 
\bea
 T_l(k',k;E)&=&\int dr u_{k'l}(r)V(r)u_{kl}(r) \\
         &&+\int dr u_{k'l}(r)V(r)\frac{1}{E+i\epsilon-H_l}V(r)u_{kl}(r)
\eea
and
\be 
u_{kl}(r)=rj_l(kr)\,,\;\;
   H_l=-\frac{1}{2m}\frac{d^2}{dr^2}+\frac{l(l+1)}{2mr^2}+V(r)\,. 
\ee
The partial wave contribution $T_l$ is related to the phase 
shift $\delta_l$ by
\be 
T_l(k,k;k^2/2m)=-\frac{1}{2mk}e^{i\delta_l}\sin\delta_l\;\;\mbox{or}\;\;
     \tan\delta_l=\frac{{\cal I}m T_l}{{\cal R}e T_l}\,. 
\ee
For S-wave scattering of a particle of mass
$m$ from $V(r)=-e^{-r/a}/ma^2$ \\
the exact solution is~\cite{MorseFeshbach}
\be 
\delta_0(k)=\mbox{Arg}(J_{2ika}(2\sqrt{2})\Gamma(1+2ka))-2ka\ln\sqrt{2}\,. 
\ee

To solve this problem numerically,~\footnote{The method presented
here is not by any means the best way to solve this particular
problem.  It is introduced purely for the purpose of testing the
recursion method.}  we introduce a discrete 
approximation with $M$ interior points between $r=0$ and
a radial cutoff $R$.  The step size is
\be 
\Delta r=\frac{R}{M+1}\,. 
\ee
Integrals are approximated by the trapezoidal rule,
and derivative operators by finite differences.
The radial cutoff is kept larger than $2\pi/k$ in order
to include one complete wavelength at the chosen energy.
The terminator for the recursion is generated from
the second Born approximation
\be
T_l^{(2)}\simeq\int dr dr' u_{k'l}(r')V(r')g_{El}^{(0)}(r',r)V(r)u_{kl}(r)\,, 
\ee
where
\be  
g_{El}^{(0)}(r',r)=-\frac{mi\kappa}{2\pi}
                       rr'j_l(\kappa r_<)h_l^+(\kappa r_>)
\ee
and $\kappa=\sqrt{2mE}$.

\begin{figure}[t]
\centerline{\psfig{figure=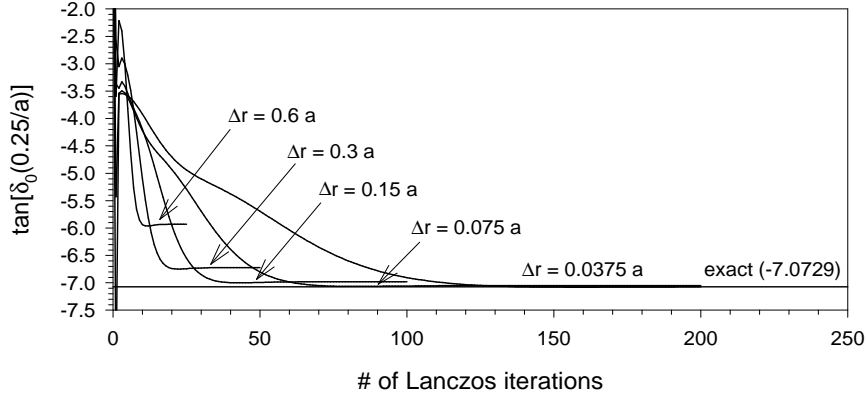,width=4.7in}}
\caption{%
Phase shift for S-wave scattering
from an exponential potential, as computed by the recursion
method with termination based on the second Born approximation.  
$\Delta r$ is the spacing used in
trapezoidal approximations to the integral expression for the partial-wave
$T$ matrix element.
\label{fig:NonRelExample}}
\end{figure}

Figure~\ref{fig:NonRelExample} displays the results of calculations 
done for $k=0.25/a$, $R=30a>2\pi/k\simeq 25a$, and $M=49$, 99, 199,
399, and 799.  They show good convergence properties.  More importantly,
they show that the imaginary part of the $T$-matrix element is
calculated reliably.  

\section{Scattering of Composites} \label{sec:Composites}

The formulation in Sec.~\ref{sec:CrossSections} assumes
implicitly that the particles involved in the scattering
process are elementary.  In a field-theoretic setting
this is not typically the case for a nonperturbative
calculation.  The particles $A$, $B$, $C$, and $D$ can
be themselves complicated bound states of the full
Hamiltonian.  It is then difficult to determine the
nature of the interaction $V$ that appears in the $T$ operator.

To construct the form of the $T$ operator in this case,
we generalize a construction given by Wick~\cite{Wick}
to compute the $S$ matrix
\be 
S_{AB\rightarrow CD}=\langle CD-|AB+\rangle 
\ee
from scattering eigenstates of the light-cone Hamiltonian
$H_{\rm LC}$.  Let $A^\dagger$ and $B^\dagger$ create 
one-particle solutions from the Fock vacuum $|0\rangle$, 
so that 
\be
|A\rangle=A^\dagger|0\rangle\,,\;\; 
H_{\rm LC}|A\rangle=\frac{m_A^2+p_{A\perp}^2}{p_A^+/P^+}|A\rangle\,.
\ee
The two-particle scattering eigenstates of $H_{\rm LC}$ are then
\be 
|AB\pm\rangle=B^\dagger|A\rangle
  +\frac{1}{s_{AB}\pm i\epsilon-H_{\rm LC}}V_B|A\rangle\,, 
\ee
where 
\be
V_B\equiv [H_{\rm LC},B^\dagger]
          -\frac{m_B^2+p_{B\perp}^2}{p_B^+/P^+}B^\dagger
\ee
and
\be
s_{AB}=\frac{m_A^2+p_{A\perp}^2}{p_A^+/P^+}+
                 \frac{m_B^2+p_{B\perp}^2}{p_B^+/P^+}
\ee
is the invariant mass of the scattering state and
the Mandelstam $s$ of the scattering process.
The definition of $V_B$ is chosen to ensure that
\be
H_{\rm LC}|AB\pm\rangle=s_{AB}|AB\pm\rangle\,.
\ee

The $T$ operator is constructed by reducing the
$S$ matrix to 
\be 
S_{AB\rightarrow CD}
    =\delta_{CD,AB}-2\pi i\delta(s_{AB}-s_{CD})T_{{\rm LC}fi}\,. 
\ee
We find
\be
T_{{\rm LC}fi}=P^+T_{fi}=\langle CD-|V_B|A\rangle\,, 
\ee
\clearpage \noindent
which can be expanded as
\be  \label{eq:TLC}
P^+T_{fi}=\langle C|V_D^\dagger
            \frac{1}{s_{AB}+i\epsilon-H_{\rm LC}}V_B|A\rangle
                     +\langle C|DV_B|A\rangle\,,
\ee
with
\bea \label{eq:2ndTerm}
\langle C|DV_B|A\rangle&=& 
       \langle C|V_B\frac{1}{s_{AB}-\frac{m_B^2+p_{B\perp}^2}{p_B^+/P^+}
                             -\frac{m_D^2+p_{D\perp}^2}{p_D^+/P^+}
                               -H_{\rm LC}}V_D^\dagger|A\rangle 
\nonumber \\
    && +\langle C|[D,V_B]|A\rangle\,. 
\eea
The first term in (\ref{eq:TLC}) can be interpreted as a direct scattering
of $B$ by $A$.  The first term in (\ref{eq:2ndTerm}) is
the crossed contribution, where $D$ is produced before $B$ is
absorbed.  The second term in (\ref{eq:2ndTerm}) is
a generalized seagull contribution, which did not appear
in Wick's formalism.~\cite{Wick}

To use these expressions efficiently, $\langle C|DV_B|A\rangle$
should be computed explicitly rather than work with the 
decomposition into crossed and seagull terms.  This is because
the decomposition introduces an inverse of the Hamiltonian.
The inverse is unavoidable for the direct term, and here
one would apply the recursion method discussed in the previous
sections.  For both terms one needs the action of $V_B$ on
the one-particle state $|A\rangle$.

\section{Concluding Remarks} \label{sec:Conclude}

A method for the nonperturbative calculation of cross sections
has been presented in the context of light-cone quantized
field theories.  The key expression is (\ref{eq:TLC}), where
the light-cone $T$ matrix is given in terms of known operators
and composite-particle eigenstates.  The terminated recursion
technique of Haydock~\cite{Haydock} is to be used to handle the
inverse of the light-cone Hamiltonian.  Work on an application
to a simple model~\cite{PV} is in progress.  One direction to
consider for possible improvements to the method is to explore
variational formulations analogous to the Schwinger and Kohn
variational principles~\cite{variational} used in nonrelativistic
scattering.

\section*{Acknowledgments}

The work reported here was supported
in part by the Department of Energy,
contract DE-FG02-98ER41087.

\end{document}